# Crystal forming ability of amorphous refractory metals under nanoindentation: a molecular dynamics study

P. Dwivedi[1], A. Fraile[2], T. Polcar[1,3]

[1]Department of Control Engineering, Faculty of Electrical Engineering, Czech Technical University in Prague, Technická 2, 166 27 Prague 6, Czech Republic

[2]Catalan Institute of Nanoscience and Nanotechnology (ICN2), BIST & CSIC Campus UAB, Bellaterra, Barcelona, Spain

[3]Engineering Materials, Department of Mechanical Engineering, University of Southampton, Southampton SO17 1BJ, United Kingdom

**Abstract**

Amorphous refractory metal coatings combine high hardness with chemical inertness, yet their metastability makes them prone to mechanically induced crystallisation (devitrification) under contact loading, and how readily such a glass re-orders to its parent body centred cubic (bcc) crystal is unknown across the refractory series. We prepared amorphous V, Nb, Mo, Ta and W by melt quenching to 300 K, benchmarked each interatomic potential against ab initio liquid radial distribution functions and self-diffusion coefficients, and probed them by large scale molecular dynamics nanoindentation. Indentation drives a localised amorphous to bcc transformation by bulk nucleation, growth and coalescence. A crystal forming ability (CFA), the maximum slope of the sigmoidal bcc fraction versus depth curve, spans about a factor of four and decreases as V > Mo > Nb > Ta > W; it falls with indenter velocity as CFA $\propto v^{-m}$ with a mean exponent of 1.08, so CFA·v is nearly constant and the transformation rate is almost velocity independent. Transforming atoms carry excess non affine displacement and local shear strain, not hydrostatic pressure, marking a shear associated displacive pathway; the bulk driving force is largest for the most resistant element, W, so resistance tracks cohesive bond strength rather than the thermodynamic driving force. Early bcc like order correlates with the sharpness and depth of the transition, whereas the mechanical work to half transformation correlates with the peak density of persistent bcc clusters. A cluster population balance links nucleation, growth and coalescence to a terminal microstructure whose completeness does not follow the CFA order.

## 1. Introduction

Metallic glasses retain the topological disorder of the liquid in a solid that reaches strengths approaching the theoretical shear limit and sustains large elastic strains [1,2]. Almost all glass forming metallic systems are multicomponent. Atomic size mismatch, negative enthalpies of mixing and deep eutectics depress the liquidus and frustrate the nucleation and growth of any single competing crystal, which permits vitrification at the cooling rates available in conventional casting [2,3]. The resulting solid is not structurally featureless: efficiently packed, solute centred clusters and fivefold symmetric motifs define a short and medium range order that is geometrically incompatible with long range translational periodicity [4].

Pure transition metals possess none of this chemical frustration. Their high crystal symmetry and rapid growth kinetics make the body centred cubic (bcc) phase easy to nucleate, and retaining a liquid like configuration to room temperature was long considered impractical. Zhong *et al.* showed that elemental refractory metals can nonetheless be vitrified by extracting heat into an adjacent crystalline sink on a nanosecond timescale, at effective rates of order $10^{14}$ K $s^{-1}$ [5]. These rates lie well beyond melt spinning or splat quenching, but they coincide with the rates imposed routinely in molecular dynamics (MD) melt quench protocols. MD therefore reaches the required cooling rate, resolves the atomic configuration directly, and carries a single interatomic model through melting, vitrification and subsequent deformation. Chemical simplicity does not imply simple crystallisation kinetics: competing crystal like and icosahedral preordering can alter the transformation pathway and change crystallisation rates between single element systems by more than two orders of magnitude [6].

Plasticity in a glass is not carried by dislocations. Strain is accommodated instead by localised, cooperative atomic rearrangements, described as flow defects or shear transformation zones (STZs) [3,7,8,9], whose characteristic volumes have been extracted from rate dependent yielding measurements [10]. Instrumented and simulated indentation impose the inhomogeneous, multiaxial stress state in which STZ activity, shear localisation and structural change compete, which is why indentation has become a standard probe of these processes [11,12].

Because the amorphous phase is metastable with respect to the bcc crystal, it retains a driving force for solid state transformation that can be released mechanically as well as thermally. Nanocrystals have been resolved within shear bands of amorphous alloys after severe deformation [13] and beneath room temperature indents in Zr based bulk metallic glasses [14,15]. The extent of transformation depends on the imposed strain field: cross sectional microscopy showed more extensive nanocrystallisation beneath a cube corner indenter than beneath a Berkovich indenter at equal peak load [16], and spatially resolved analysis of Vickers indents related the degree of crystallisation to the local stress state [17]. The effect is not universal. Synchrotron analysis of the deformation zone beneath cylindrical indents in a Zr based glass detected no crystallisation [18], so indentation induced devitrification depends on the alloy, the indenter geometry and the detection limit rather than following automatically from contact loading. Contact driven phase change is separately well established in covalent semiconductors, where the stress field beneath an indenter drives polymorphic and crystalline to amorphous transitions [19,20].

Atomistic simulation resolves the sub indenter region that experiment reaches only after sectioning. Shi and Falk showed that the localisation pattern and short range structural response of a simulated metallic glass during indentation depend on both preparation history and indentation rate [21]. Later work established that heterogeneous rearrangement continues during unloading and accumulates under repeated contact, so retraction cannot be treated as a structurally inactive reversal of the loading path [22,23,24,25]. Uniaxial compression of amorphous Ni produced fcc, bcc and hcp populations in sequence, showing that mechanical loading can drive multistage ordering in a single element glass [26]. The closest precedent for the present material class is the work of Cao et al., who vitrified Ta by MD melt quenching and observed progressive bcc ordering under elastic cycling, including in a fully periodic cell without a free surface or crack, while a time matched thermal calculation did not crystallise [27]. That study concerns one element, one empirical potential and a uniaxial cyclic path rather than contact loading. Deformation has also been used deliberately to select phases: Yavas et al. reported amorphous to bcc conversion in Nb layers and amorphous to hcp conversion in Zr layers during indentation of Zr/Nb nanolayers, with the transformed phases stabilised by the heterophase interfaces [28]. The configurations studied here contain no crystalline seed layers, interfaces or multilayer confinement, so that interface assisted stabilisation is not available.

The resistance of a melt to crystallisation on cooling, its glass forming ability (GFA), is described by a mature set of thermal descriptors, among them the reduced glass transition temperature and the width of the supercooled liquid region [29,30], which build on classical nucleation [31] and free volume transport [32] descriptions of the supercooled melt. No comparable normalised descriptor exists for the inverse propensity: the ease with which a deformed amorphous metal returns to its crystalline ground state under mechanical work. Thermal GFA criteria are in any case not transferable to a stress driven transition below the glass transition temperature, where local shear work and non hydrostatic strain, rather than thermal fluctuation alone, supply the bias. How this propensity varies across the bcc refractory metals has not been established. A meaningful comparison requires a common structural metric, loading geometry, temperature and analysis procedure, together with an assessment of each interatomic potential in the liquid state, because the quenched glass inherits its short and medium range topology from the simulated melt, and agreement with 0 K lattice parameters and elastic constants alone does not establish transferability to melt quench and devitrification simulations.

The present study addresses this gap. Amorphous V, Nb, Mo, Ta and W are prepared by MD melt quenching, each interatomic potential is benchmarked against published ab initio liquid structure and self-diffusion data, and the glasses are subjected to spherical nanoindentation at 300 K under a common loading geometry and analysis procedure. The bcc fraction is followed during loading and converted into a depth normalised crystal forming ability (CFA), defined operationally as the maximum slope of a logistic fit to the bcc fraction against indentation depth. CFA measures the transformation produced per unit indenter advance under the stated protocol; it is deliberately not identified with a thermodynamic nucleation rate, an intrinsic material constant, or the inverse of GFA, because indentation couples elapsed time, hydrostatic pressure, deviatoric strain, local heating and structural relaxation, and a correlation with any single one of these does not by itself isolate a causal contribution. On this common basis we ask how the propensity for contact driven devitrification varies across the five bcc refractory metals, and which structural and mechanical variables govern the transformation pathway.

## 2. Computational methods

### 2.1. Interatomic potentials

All MD simulations were performed with Large-scale Atomic/Molecular Massively Parallel Simulator (LAMMPS) [33,34] in metal units. Newton’s equations were integrated with velocity Verlet using a 0.1 fs timestep. For each refractory metal, the selected potential reproduced the bcc lattice parameter, cohesive energy and bulk modulus against density functional theory (DFT) or experiment (Table 1). Because the simulations include melting and quenching, preference was given to potentials whose fitting or validation databases sampled high temperature or liquid configurations, rather than only the 0 K crystal.

For V we used the Olsson embedded atom method (EAM) [35] potential [36], which reproduces the bcc elastic and lattice stability data relevant to configurations far from the ground state minimum. For Nb we used the force matched EAM potential of Fellinger, Park and Wilkins [37], whose database includes high temperature bcc and liquid structures. For Mo we used the Mo branch of the Starikov *et al.* U-Mo angular-dependent potential (ADP) [38]; its dipole and quadrupole terms extend the EAM description of directional d bonding [39]. For Ta we used the Chen *et al.* W-Ta Finnis-Sinclair (FS) [40] potential [41], which was validated against two phase melting and reproduces the bcc Cauchy violation. For W we used the Marinica *et al.* EAM2 potential [42], whose database includes ab initio liquid forces.

Table 1 compares the equilibrium lattice parameter $a_0$, cohesive energy magnitude $E_{coh}$ and bulk modulus B from static energy volume and DFT or experimental references [43,44]. Values of this kind depend on the potential and on the calculation methodology, and that variability has been quantified across large potential databases [45]. The quantity $\gamma_s$ listed in Table 1 is a deliberately crude broken bond energy scale, $\gamma_s = E_{coh}/(12\ a_0^2)$, used only as a per area cohesion descriptor in the downstream scaling analysis.

| **Quantity** | **V** | **Nb** | **Mo** | **Ta** | **W** |
|---|---|---|---|---|---|
| $a_0$ (Å), this work | 3.030 | 3.308 | 3.149 | 3.304 | 3.140 |
| $a_0$ (Å), exp. | 3.03 | 3.30 | 3.15 | 3.30 | 3.16 |
| $E_{coh}$ (eV/at.), this work | 5.31 | 7.09 | 6.88 | 8.10 | 8.90 |
| $E_{coh}$ (eV/at.), exp. | 5.31 | 7.57 | 6.82 | 8.10 | 8.90 |
| B (GPa), this work | 157.0 | 172.0 | 303.3 | 194.2 | 320.0 |
| B (GPa), exp. | 162 | 170 | 272 | 200 | 323 |
| Coexistence $T_m$ (K) | 2138 | 2685 | 2770 | 3355 | 3874 |
| $T_m$ (K) exp. | 2183 | 2750 | 2896 | 3293 | 3695 |
| ρ (g $cm^{-3}$) exp. | 6.081 | 8.521 | 10.206 | 16.664 | 19.720 |
| $\gamma_s$ (J $m^{-2}$) | 0.77 | 0.87 | 0.93 | 0.99 | 1.21 |

**Table 1.** Crystalline reference state benchmark of the five selected potentials. Values labelled “this work” are from static energy volume scans; experimental values are from Kittel [44] and the DFT or experimental compilation of Derlet *et al.* [43], with methodological variability as discussed in Ref. [45]. Coexistence $T_m$ are two-phase solid-liquid melting temperatures of each potential, $\gamma_s$ is the broken bond cohesion per area descriptor defined in the text.

### 2.2. Melt quench amorphisation

For each element, three independent amorphous configurations were generated by repeating the complete melt quench protocol using distinct random seeds. Each preparation started from a body-centred cubic crystal in a simulation cell of approximately 400 × 400 × 210 Å³, containing approximately $1.9 \times 10^6$ to $2.4 \times 10^6$ atoms, depending on the lattice parameter. Melting and quenching were performed with periodic boundaries in the x, y, and z directions. Each independently equilibrated liquid was quenched separately to produce three glass replicas per element and 15 glass configurations in total. Before indentation, each equilibrated glass was converted to a slab with periodic boundaries in the x and y directions, a nonperiodic z direction, a free top surface, and a fixed 10 Å bottom layer. Because the glass is less dense than the bcc crystal, the relaxed slabs are slightly larger than the initial cell, the quenched cells measure $L_x \approx L_y \approx$ 40.3-41.0 nm and $L_z \approx$ 21.1-21.4 nm, 1.1-4.3% larger in volume than the initial bcc cells; $L_z$ is the thickness H of the indentation slab. The fixed layer comprises the atoms with z ≤ 10 Å, i.e. ≈1 nm and ≈4.7% of the atoms (SM, Table S8). The liquids were equilibrated using anisotropic barostatting and quenched to 300 K under zero pressure isothermal-isobaric ensemble (NPT) control (Figure 1). The liquid superheats were $T_{max}/T_{m,exp}$ = 1.10 to 1.82, and the principal quench rates were 1.4 to 4.6 × $10^{13}$ K $s^{-1}$. Each glass was accepted after its pressure, temperature, density, and potential energy became stationary. The complete preparation and acceptance procedure was applied independently to every glass replica.

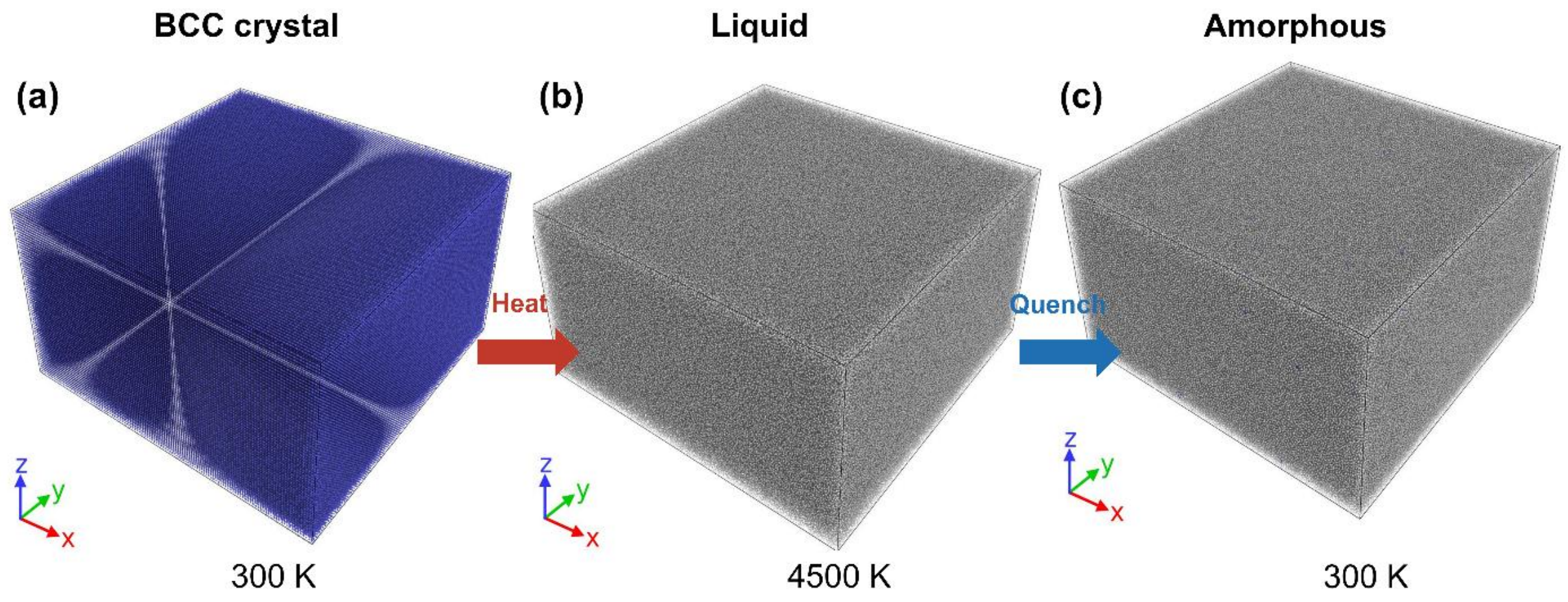

Figure 1. Melt quench amorphisation route used to prepare the monatomic refractory glasses, illustrated for a representative cell of approximately 2.4 × $10^6$ atoms and 40 × 40 × 21 $nm^3$. (a) bcc single crystal at 300 K; (b) liquid obtained by heating above the coexistence melting temperature ($T > T_m$); (c) amorphous solid recovered by quenching to 300 K at 1.4 to 4.6 × $10^{13}$ K $s^{-1}$.

### 2.3. Liquid structure validation against ab initio MD

The final configurations were first screened using the radial distribution function g(r). All showed a broad first neighbour peak, a low but non zero first minimum, a split second peak and decay to g(r) = 1 without Bragg like features. The production radial distribution functions of the five glasses and their second-peak splitting are shown in the SM (Figs S1 and S2), confirming that every sample is amorphous. These features establish the loss of long range periodicity, but they do not by themselves show that a potential reproduces the correct disordered structure.

We therefore benchmarked the liquid structure against ab initio MD. Converged ab initio molecular dynamics (AIMD) trajectories for rapidly quenched monatomic glasses are unavailable, whereas near melting liquid references exist for all five elements: V [46], Nb [47,48], Mo [48], Ta [49,50] and W [49]. This is a relevant test for amorphisation because the glass inherits its short and medium range topology from the liquid precursor.

For this comparison, separate liquids of each potential (250,000 atoms, periodic cell (50 × 50 × 50), zero pressure, Nosé-Hoover thermostat and barostat, $\Delta t = 1$ fs) were equilibrated for 20 ps at the temperature of each ab initio reference (V 2173 K; Nb 2800 K; Mo 2950 K; Ta 3350 K; W 3750 K) and sampled for a further 60 ps. Their g(r), averaged over the last 30 ps, and the ab initio curves were interpolated onto a common radial grid. Agreement was evaluated over $1.9 \le r \le 6.0$ Å using root-mean-square error (RMSE), normalised RMSE (NRMSE), Pearson *r*, first peak position and height, and second peak position. NRMSE is RMSE divided by the reference peak to baseline range. Peak positions and RMSE are the primary measures; Pearson *r* is less discriminating because it is dominated by the first maximum. Three limitations apply. The reference temperature and density are element specific, the source curves differ in grid resolution and normalisation, and comparison beyond about 6 Å, close to half the edge length of the 100-120 atom AIMD cells [48,49], is sensitive to those source differences. Because g(r) does not characterise atomic mobility, the self-diffusion coefficient D of the same liquids was obtained from the mean-squared displacement (MSD), computed over all time origins after removal of the centre-of-mass drift, through $MSD(t) = 6Dt$ fitted over 10-50 ps, with ten independent velocity seeds per state point. (SM, Section S8).

### 2.4. Nanoindentation and structural analysis

Spherical nanoindentation was performed at 300 K with a non-atomistic, frictionless, purely repulsive spherical indenter of radius R = 100 Å, implemented with the LAMMPS fix indent [34]. An atom at distance r from the indenter centre experiences the radial repulsive force $F(r) = -K(R - r)^2$ for $r < R$ and zero otherwise, with $K = 0.5$ eV Å$^{-3}$, corresponding to the energy $(K/3)(R - r)^3$; the indenter is therefore a smooth repulsive wall of finite stiffness rather than a hard sphere or an inverse-power potential, as is standard in atomistic indentation [51]. The indenter centre lies at x = y = 200 Å, its lowest point starts at z = 211.4 Å, at the level of the free surface, and it moves along −z at constant velocity v. The indentation depth is the indenter displacement, $h = vt$; because the first non-zero indenter force is recorded at $h \le 0.17$ nm, h equals the penetration of the lowest point of the sphere below the initial free surface to within 0.17 nm. The indentation load F is the z component of the total force exerted by the indenter on the atoms, positive in compression, written every 0.1 ps and converted from eV $Å^{-1}$ to nN (1 eV $Å^{-1}$ = 1.602 nN). The mobile atoms were kept at the target temperature by velocity rescaling every 110 timesteps, and the atoms of the fixed layer were held at their positions after an initial energy minimisation. The reference velocity was 0.075 Å $ps^{-1}$; additional trajectories used 0.15 and 0.30 Å $ps^{-1}$. For each element, undeformed copies of all three independently prepared glasses were indented at each of the three velocities. Ten trajectories were initiated for each combination of glass and velocity from identical undeformed atomic coordinates using distinct random seeds. Each trajectory was analysed separately. Within each combination of glass and velocity, the trajectory level CFA values were summarized using the mean and standard deviation across the ten seeded trajectories. Local bcc environments were identified in the Open Visualization Tool (OVITO) [52] using polyhedral template matching (PTM) [53] with a root-mean-square deviation (RMSD) cutoff of 0.12. At each saved configuration, atoms classified as bcc were selected and grouped into spatially connected clusters using the Cluster Analysis modifier. Clusters containing fewer than 100 bcc classified atoms were excluded. This threshold was used as an operational filter for small local PTM motifs and was not interpreted as a critical nucleus size. Because clusters are defined by spatial connectivity, a cluster may contain more than one crystallographic orientation. The atoms of the fixed bottom layer never move, so their PTM classification at every depth is that of the undeformed glass. Future transforming atoms were compared with matched amorphous controls (Section 3.6) using the non-affine displacement $D^2_{min}$ [9], the local von Mises shear strain [54], the hydrostatic pressure and the atomic volume. For atom i with neighbours j within a cutoff $r_c$, $D^2_{min,i} = \min_J \Sigma_j |d_{ij} - J\, d^0_{ij}|^2$ is the residual of the best local affine map J between the neighbour vectors $d^0_{ij}$ of a reference configuration and $d_{ij}$ of the current configuration [9]. It vanishes for any homogeneous local strain or rigid-body motion, however large, and measures rearrangements that no single strain can describe, such as the neighbour exchanges of a shear transformation zone. The local shear strain is the von Mises invariant of the Green-Lagrangian strain tensor obtained from the same best-fit J [54]. Both were computed with the OVITO Atomic Strain modifier with $r_c$ equal to 1.2 times the bcc nearest-neighbour distance (3.15-3.44 Å) and the preceding saved configuration as reference, so that they quantify the deformation in the interval immediately before transformation. The hydrostatic pressure is the per-atom virial stress divided by the Voronoi atomic volume. For each element, one glass replica was additionally propagated at 300 K with no indenter for a time matched to the longest indentation trajectory, under identical thermostatting and boundary conditions.

### 2.5. Early order kinetic model and statistical treatment

The ensemble mean bcc fraction curve of each element (three glass replicas × ten seeds) was fitted with a depth domain form of the Kolmogorov Johnson Mehl Avrami (KJMA) equation [55,56,57],

$$f(x) = f_0 + (f_\infty - f_0)\left[1 - \exp\left(-\left(\frac{x}{h_c}\right)^{n_h}\right)\right], \quad (1)$$

$$x = h - h_1, h_1 = 0.1\, nm$$

For the primary analysis, $f_0$ was fixed to the first analysed value at $h_1$ = 0.1 nm, and the fit coordinate was shifted to $x = h - h_1$. Because the indentation field and affected volume evolve with depth, $n_h$ is an effective shape parameter rather than a classical isothermal Avrami exponent [55,56]. Here f is the bcc fraction, $f_\infty$ the terminal bcc fraction and $h_c$ a characteristic transformation depth. The half-transformation depth, at which f reaches $f_0 + (f_\infty - f_0)/2$, is $h_{50} = h_1 + h_c(\ln 2)^{1/nh}$.

## 3. Results and discussion

### 3.1. Liquid structure benchmark: classical MD versus ab initio reference

Figure 2 and Table 2 compare the classical liquids, equilibrated at the temperatures of the ab initio references (Section 2.3), with the AIMD g(r). All five potentials reproduce a broad first-neighbour maximum, a first minimum and a second-neighbour maximum without crystalline features; Pearson r ≥ 0.979. First- and second-peak positions agree within 0.13 Å. The first-peak height is reproduced within 2% for V, underestimated by 11% for Nb and 14% for Mo, and overestimated by 9% for Ta and 20% for W (a parity plot of the first-peak positions and heights is given in SM, Fig. S3). This separation of position from amplitude is expected: peak positions are set by the pair equilibrium length scale to which the potentials are fitted, whereas peak amplitudes depend on the curvature of the potential and on the temperature relative to the melting point of each model.

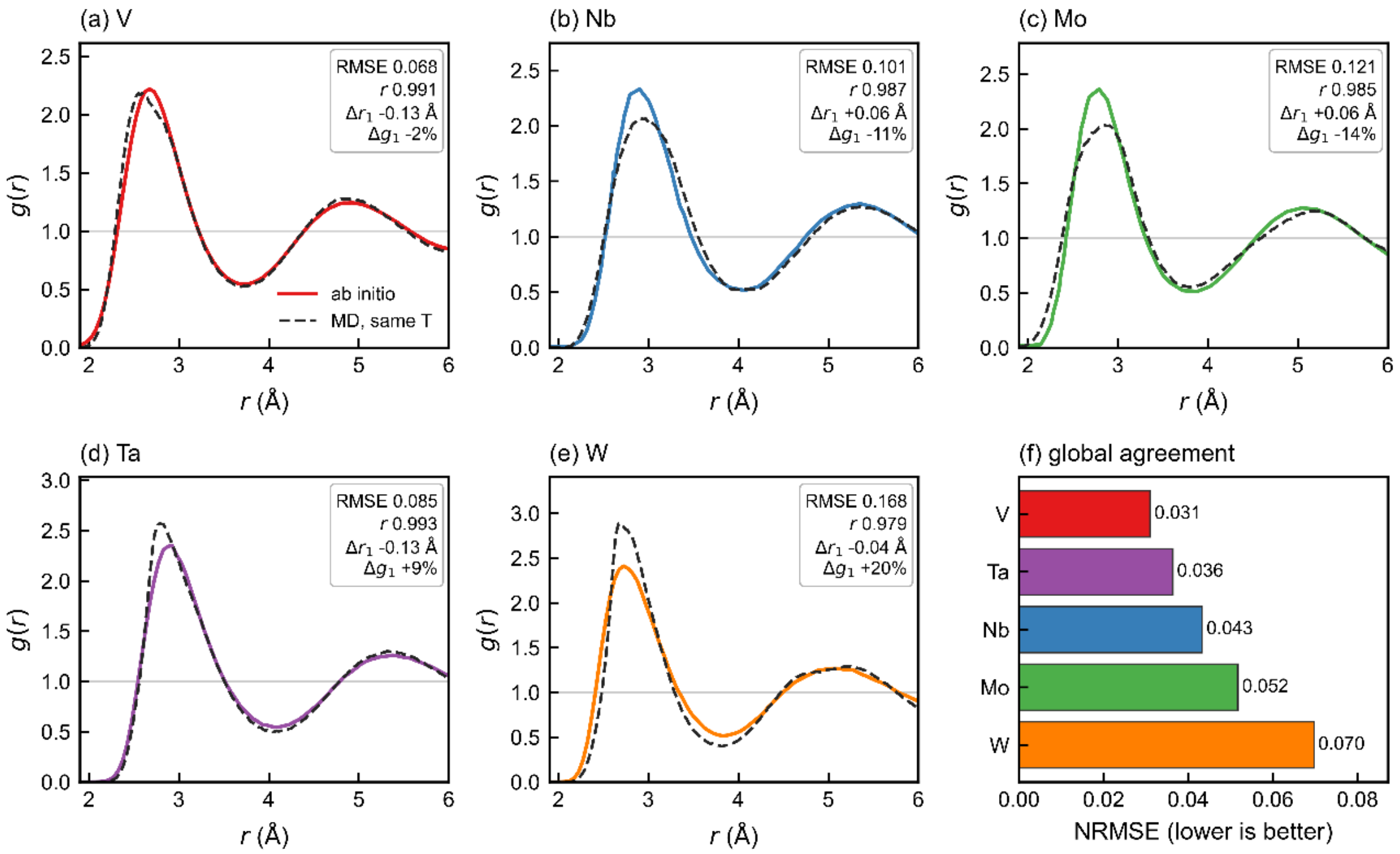

**Figure 2.** MD liquid radial distribution functions (black dashed), equilibrated at the temperatures of the ab initio references, overlaid on the AIMD g(r) curves (coloured solid) for (a) V, (b) Nb, (c) Mo, (d) Ta and (e) W over 1.9 - 6.0 Å. Each panel lists the RMSE, Pearson r and the first-peak position and height differences (MD - reference). (f) Global agreement ranked by NRMSE: V < Ta < Nb < Mo < W.

Global RMSE ranks the agreement as V (0.068) < Ta (0.085) < Nb (0.101) < Mo (0.121) < W (0.168); NRMSE gives the same order. V gives the closest overall match, with $\Delta r_1 = -0.13$ Å, $\Delta g_1 = -1.6\%$ and r = 0.991. W has the sharpest first shell $\Delta g_1 = +19.7\%$, which dominates its RMSE; its peak positions nevertheless agree within 0.13 Å. These differences are modest enough to retain all five selected potentials, but they are reported because liquid state fidelity is not uniform.

The classical melts are diffusive liquids. At the reference temperatures their MSD is linear in time (log-log slope 0.99-1.01 over 10-50 ps) and gives D = 0.565 (V), 0.484 (Nb), 0.682 (Mo), 0.331 (Ta) and 0.186 $Å^2\ ps^{-1}$ (W) (SM, Table S9 and Fig. S10), of the same order as the AIMD values of 0.387 ± 0.015 (Nb) and 0.343 ± 0.015 $Å^2\ ps^{-1}$ (Mo) [48] and 0.378 ± 0.025 (Ta) and 0.307 ± 0.015 $Å^2\ ps^{-1}$ (W) [49]; the latter were obtained with 100-120 atoms and therefore underestimate the large-system value [58]. The PTM bcc fraction of the liquids does not exceed 0.0024% and no connected bcc cluster larger than two atoms forms, so partial crystallisation or structural arrest at the liquid temperatures is excluded. The second maximum of the Mo liquid is a single smooth peak, as in the AIMD reference (Fig. 2c), whereas the W liquid shows a weak low-r shoulder near 4.9 Å (Fig. 2e). An asymmetric second peak signals polytetrahedral (icosahedral-type) local order in liquid transition metals: González and González relate the asymmetric, shouldered second peak of their AIMD S(q) for liquid W to icosahedral local order [48,49], and ab initio studies of liquid and supercooled Ta and Zr find polytetrahedral local order more complex than the purely icosahedral one [50,59].

**Table 2.** Quantitative agreement between the MD liquids and the ab initio g(r) at the temperatures of the ab initio references, recomputed from the curves over the common window. Δ values are MD − reference; $\Delta g_1$ is in percent of the reference first peak height. NRMSE is normalised by the reference peak to baseline range. The first shell class is assigned by $|\Delta g_1|$: excellent (<5%), good (5 to 10%) and moderate (>10%). Overall curve shape is close for all five elements (r ≥ 0.979).

| Elem. | Range (Å) | RMSE | NRMSE | Pearson r | $\Delta r_1$ (Å) | $\Delta g_1$ (%) |
|---|---|---|---|---|---|---|
| V | 1.90 - 6.00 | 0.068 | 0.031 | 0.991 | −0.13 | −1.6 |
| Nb | 1.90 - 6.00 | 0.101 | 0.043 | 0.987 | +0.06 | −11.2 |
| Mo | 1.90 - 6.00 | 0.121 | 0.052 | 0.985 | +0.06 | −13.8 |
| Ta | 1.90 - 6.00 | 0.085 | 0.036 | 0.993 | −0.13 | +9.3 |
| W | 1.90 - 6.00 | 0.168 | 0.070 | 0.979 | −0.04 | +19.7 |

Overall, the benchmark supports the use of the selected potentials for generating amorphous configurations from well-defined liquid precursors: the nearest-neighbour length scale is reproduced for all five elements and all five classical melts are diffusive liquids. The comparison benchmarks the structure and mobility of the liquid precursor; it does not validate the free energies of the glasses or the transformation kinetics.

### 3.2. Mechanical response and potential energy relaxation

Figure 3 compares indentation load, bcc classified atom fraction, and corrected per atom potential energy for a representative V trajectory. From the first stored configuration at $h = 0.375$ nm, the potential energy decreases from -5.1231 to a minimum of -5.2053 eV atom$^{-1}$ at $h = 4.125$ nm. At the final stored depth of 9.75 nm, it remains 0.0784 eV atom$^{-1}$ below the first recorded value. The main energy decrease coincides with an increase in the bcc fraction from 4% to approximately 80%. The energy evolution is consistent with structural relaxation during indentation induced ordering.

In the time matched calculation without an indenter, the bcc fraction remained at its initial value, indicating the transformation is associated with contact loading rather than thermal evolution.

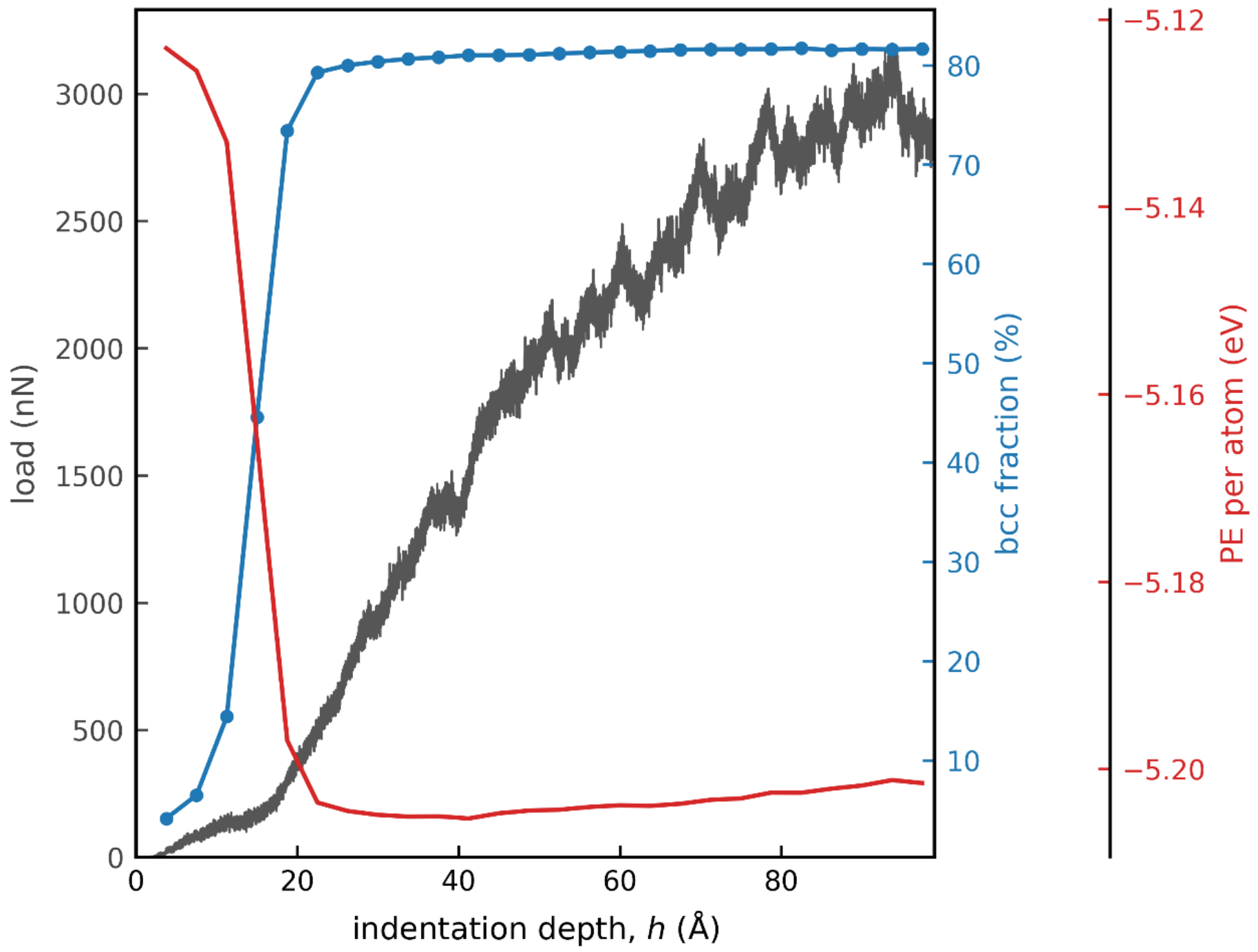

**Figure 3.** Indentation load F (grey), bcc fraction (blue) and per-atom potential energy (red) versus indentation depth h for a representative V trajectory (v = 0.075 Å ps$^{-1}$). F is the z component of the total indenter force on the specimen.

### 3.3. Rate dependent transformation kinetics and the CFA definition

Figure 4 shows the bcc classified atom fraction, for V at the three indenter velocities. Ten seeded trajectories were obtained from each of the three independently quenched glasses at every velocity. The trajectories were first averaged within each glass and then across the three glasses. Increasing velocity shifted the transformation towards greater indentation depth, broadened the transition, and slightly reduced the bcc fraction attained within the analysed depth interval. This behaviour is consistent with the shorter elapsed time available for nucleation and growth at a given indentation depth.

Each trajectory was fitted using the logistic function

$$f_{\mathrm{BCC}}(h) = S_1 + \frac{S_2 - S_1}{1+\exp[-(h-h_0)/\Delta x]} \quad (2)$$

where $S_1$ and $S_2$ are the fitted lower and upper asymptotes, $h_0$ is the inflection depth, and $\Delta x > 0$ is the transition width parameter. The operational crystal forming ability, CFA, was defined as the maximum slope of the fitted curve:

$$\mathrm{CFA} = \frac{\mathrm{d}f_{\mathrm{BCC}}}{\mathrm{d}h}\bigg|_{h=h_0} = \frac{S_2 - S_1}{4\Delta x} \tag{3}$$

Because $f_{\mathrm{BCC}}$ was fitted as a fraction between 0 and 1, CFA is reported as 100 x CFA, in percentage points $Å^{-1}$. CFA therefore represents the maximum change in the bcc classified atom fraction per unit indentation depth. $S_1$ is the lower asymptote of Eq. (2), reached only for $h \to -\infty$, outside the physical range $h \geq 0$; it is a fitting parameter, not a transformed fraction, and nothing in the logistic form requires it to be negative. Both asymptotes were constrained to $0 \leq S_1, S_2 \leq 1$, and the fitted $S_1$ is close to the small PTM-bcc background of the undeformed glass. By Eq. (3), an error $\delta S_1$ changes CFA only by the fraction $\delta S_1/(S_2 - S_1)$. Refitting with $S_1$ unconstrained or fixed at the first measured $f_{\mathrm{BCC}}$, or taking CFA model-free as the maximum of the smoothed numerical derivative of $f_{\mathrm{BCC}}(h)$, changes CFA by −4.3% to +4.9% and leaves the ranking unchanged (SM, Section S9 and Table S10).

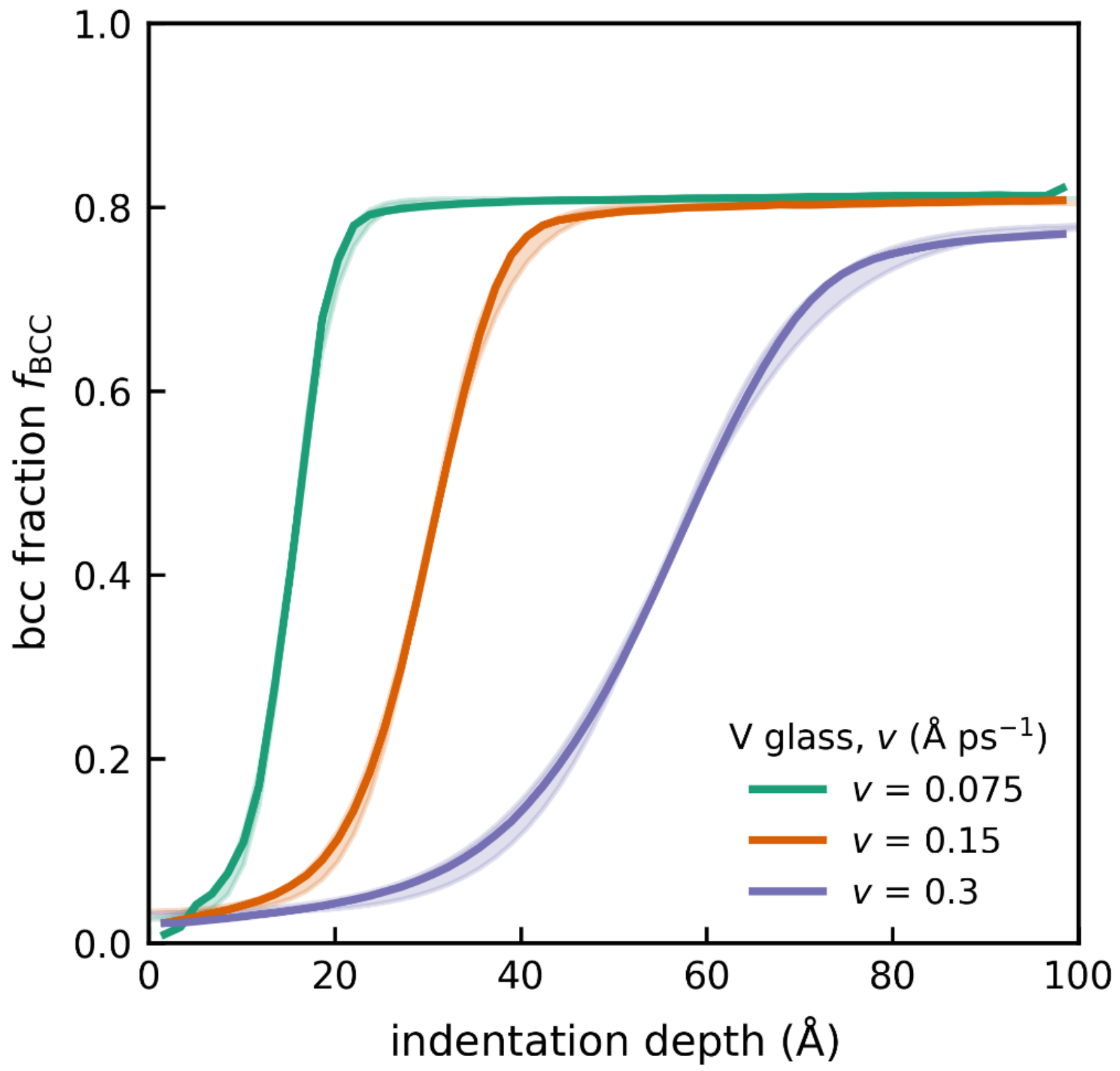

**Figure 4.** Bcc classified atom fraction versus indentation depth for V at 300 K and indenter velocities of 0.075, 0.15, and 0.30 Å $ps^{-1}$. Solid curves show the mean of the three glass specific seed means. Shaded bands represent standard deviation.

### 3.4. Crystal forming ability and its dependence on indenter velocity

Spherical nanoindentation of the glasses produces a localised, persistent amorphous to bcc transformation in every element. At 300 K and v = 0.075 Å $ps^{-1}$ reported CFA ranking is V (9.5) > Mo (8.4) > Nb (3.6) > Ta (2.9) > W (2.3), in units of 100 × CFA ($Å^{-1}$). W therefore requires the greatest indentation advance per unit transformed fraction within this series; this describes transformation resistance and not hardness.

Fitting the reported three point CFA (Figure 5) summaries gives element slopes of 0.99 (V), 1.18 (Nb), 0.97 (Mo), 1.14 (Ta) and 1.10 (W). Their arithmetic mean is 1.076 and their population standard deviation is 0.083. This is between element dispersion of five descriptive slopes, not uncertainty on a material independent exponent.

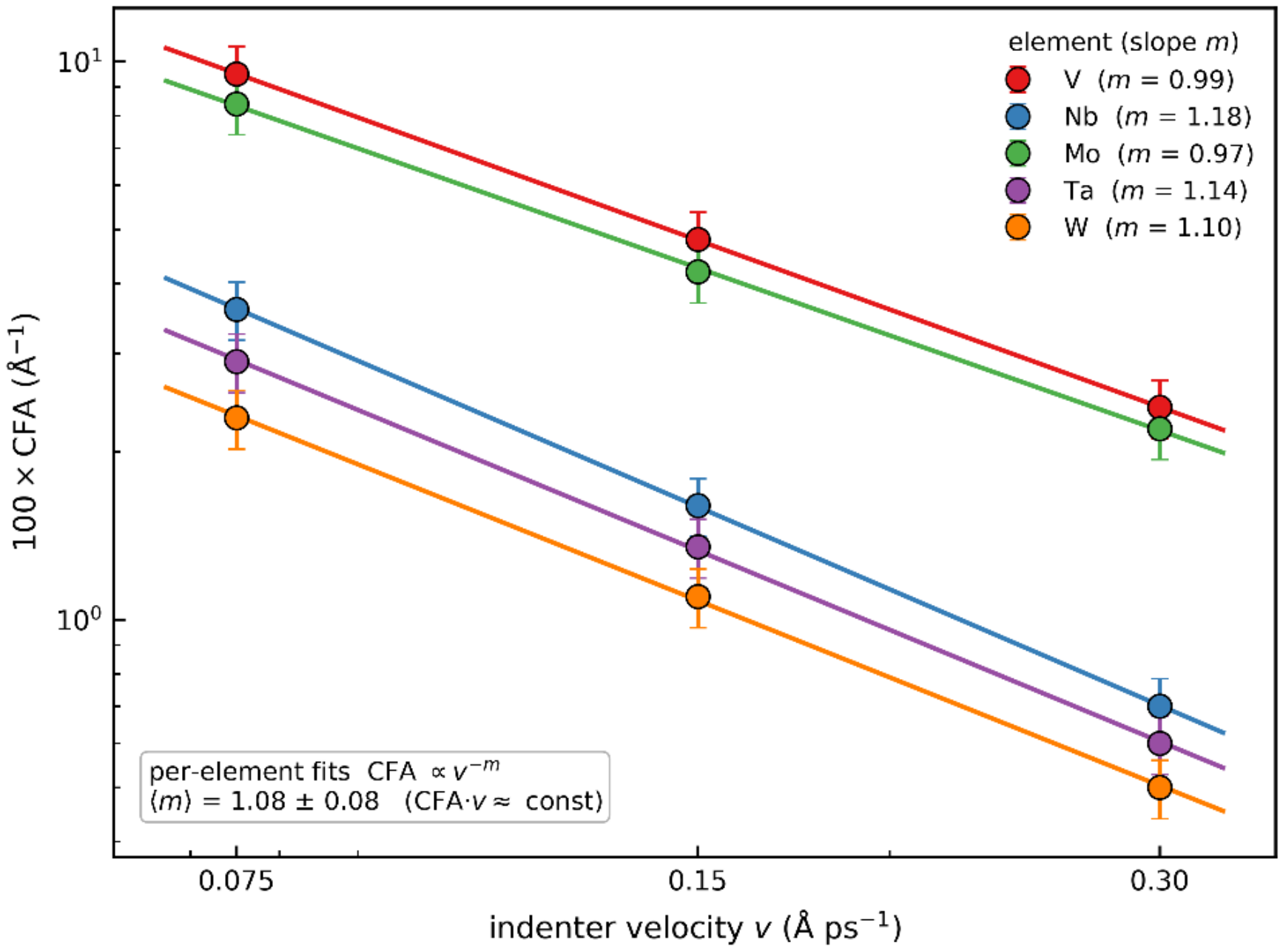

**Figure 5.** Reported crystal forming ability versus nominal indenter velocity. Lines are power law fits. The quoted 1.08 ± 0.08 is the mean ± population SD across five element specific slopes.

### 3.5. Nucleation, growth and coalescence

The atomistic configurations resolve the transformation pathway directly. Figure 6 shows the bcc classified population in amorphous Nb, compact bcc clusters first appear at several sites within the sub tip analysis volume, grow and impinge as loading continues.

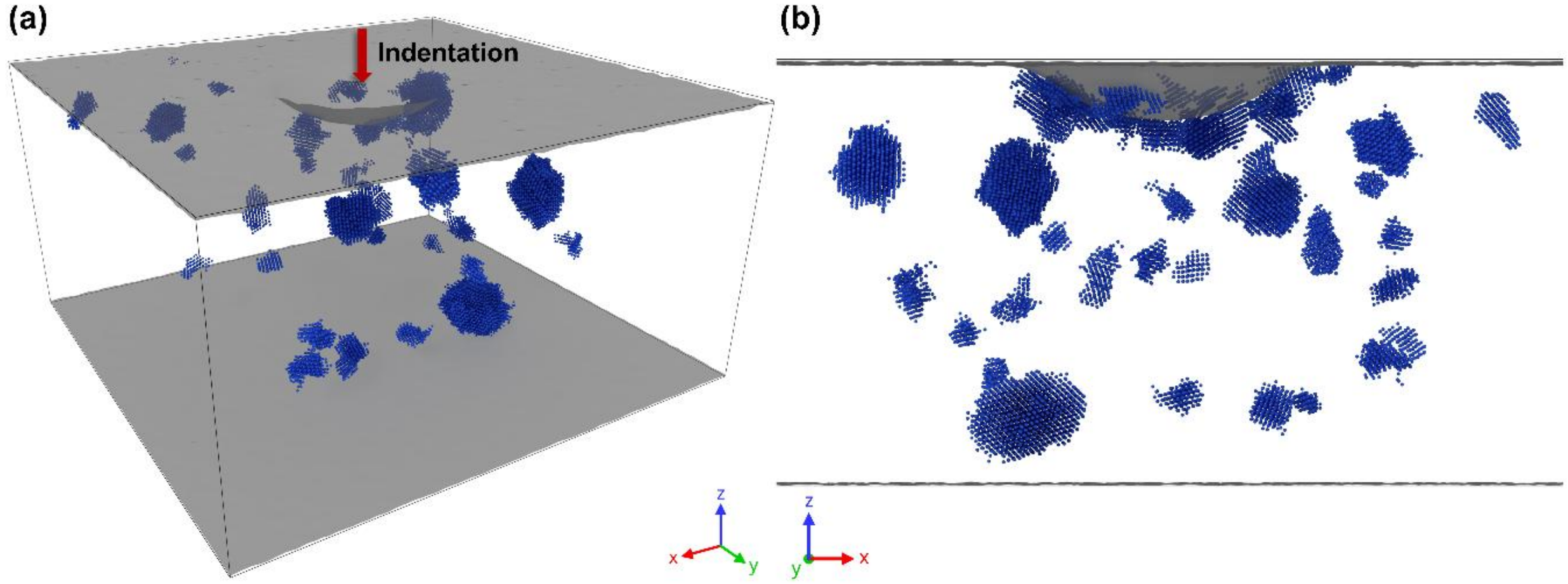

**Figure 6.** Indentation-induced devitrification of amorphous Nb: (a) perspective view and (b) side view. Only bcc-classified atoms are shown (blue); the amorphous atoms are hidden. The semi-transparent grey sheets are surface meshes that mark the free top surface with the indent and the bottom boundary of the slab. Small bcc clusters form at multiple sites within the mechanically affected volume, including regions away from the tip axis, and grow, impinge and coalesce with continued loading.

Figure 7 follows the Nb transformation from 2.5 to 10 nm. The number of retained bcc clusters rises as persistent clusters appear, reaches a maximum and then falls through impingement and coalescence, while the mean cluster volume continues to increase. Mo and V form the largest persistent populations and show the strongest later coalescence. The cluster count and mean cluster size statistics underlying this description are given in the SM (Fig. S4).

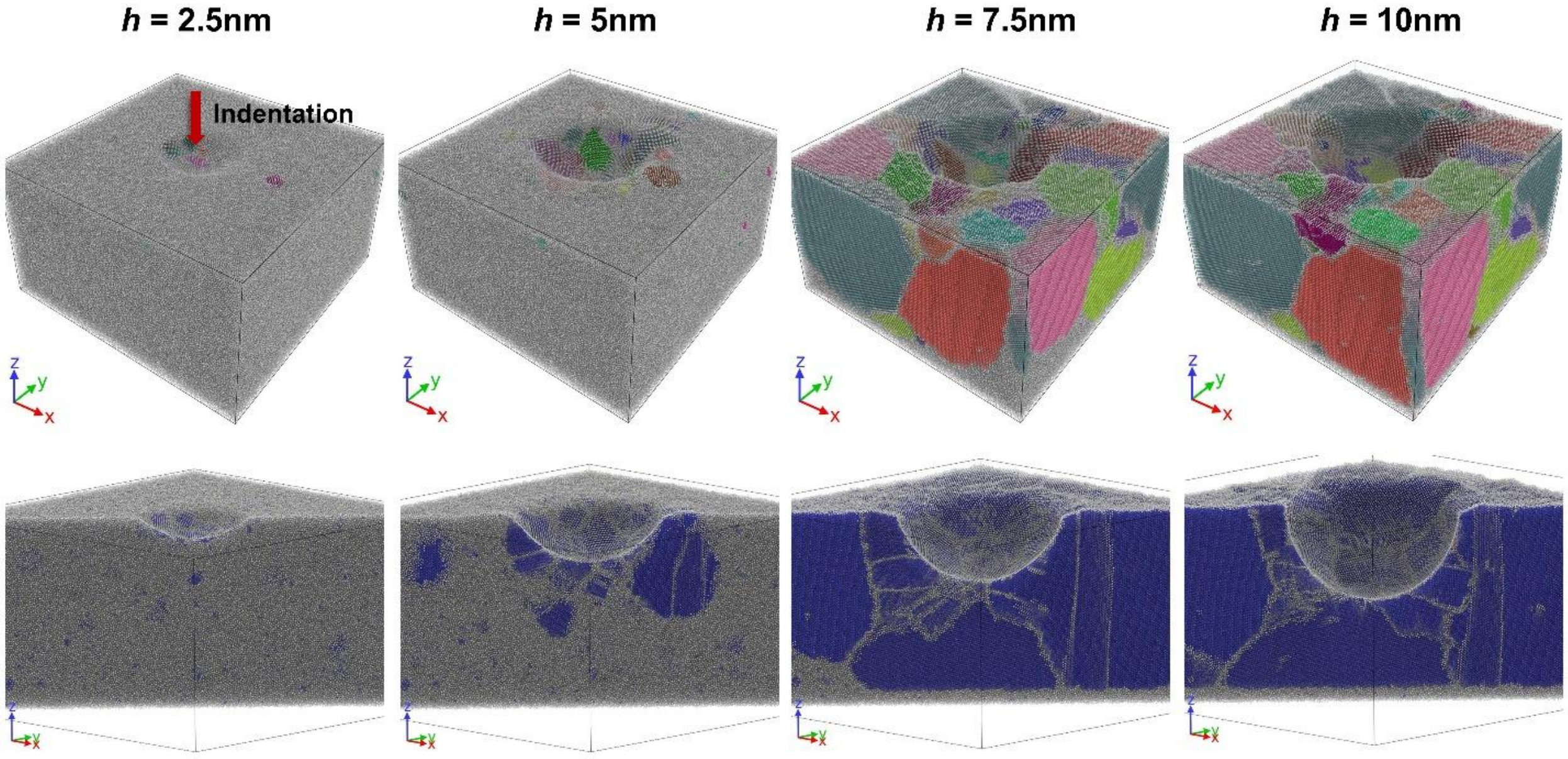

**Figure 7.** Progressive devitrification of amorphous Nb with increasing indentation depth, h = 2.5, 5.0, 7.5 and 10 nm. Top row: perspective view of the specimen with bcc grains coloured by crystallographic orientation, showing an expanding polycrystalline region around the contact. Bottom row: central y-z section showing bcc-classified atoms (blue) against the amorphous matrix (grey). Crystallisation initiates as isolated nuclei beneath the contact and develops into an extended, grain-boundary-separated polycrystalline zone that, at the largest depths, extends to the lateral periodic boundaries and approaches the fixed bottom layer, illustrating bulk nucleation, growth and coalescence under the advancing indenter.

### 3.6. Transformation is associated with enhanced local shear strain

To examine local precursors separately from the CFA analysis, we performed a time-lagged, matched case-control analysis. Atoms that were not bcc in a precursor configuration, became bcc in the next saved configuration and remained bcc in the following one (persistent transformation) were compared with control atoms that remained non-bcc in both later configurations. Controls were drawn at random, in equal number, from the same joint stratum of radial distance from the indenter axis (8 bins), height (8 bins) and hydrostatic pressure (sextiles) as the transforming atoms, and all precursor values were taken in the precursor configuration. For each precursor variable $x$, Cohen's $d$ [60] was calculated as

$$d_x = \frac{\bar{x}_{\text{transform}} - \bar{x}_{\text{control}}}{s_{p,x}} \quad (4)$$

where $\bar{x}_{\text{transform}}$ and $\bar{x}_{\text{control}}$ are the group means and $s_{p,x} = \{[(n_T - 1)s^2_{T,x} + (n_C - 1)s^2_{C,x}]/(n_T + n_C - 2)\}^{1/2}$ is the pooled within group standard deviation of the $n_T$ transforming and $n_C$ control atoms, with $s^2_{T,x}$ and $s^2_{C,x}$ the sample variances [60]. Because neighbouring atoms are spatially correlated, d is used as a descriptive standardised effect size rather than as a significance test.

For Mo, both $D^2_{\text{min}}$ and local shear strain exhibited positive $d$ over most of the indentation trajectory. Hydrostatic pressure produced only a small negative contrast, which is expected in part because the controls are matched in pressure, whereas atomic volume exhibited consistently negative $d$, indicating a smaller precursor atomic volume among future transforming atoms. In the cross element analysis, atoms that were amorphous in each precursor configuration were ranked by $D^2_{\text{min}}$ and divided into ten equal population bins. Figure 8 shows for all five elements, the persistent transformation probability increased monotonically with $D^2_{\text{min}}$, rising by approximately one to two orders of magnitude between the lowest and highest deciles. The element averaged $d$ for local shear strain was equal to or greater than that for $D^2_{\text{min}}$. Thus, in this univariate comparison, local shear strain discriminated future transforming atoms at least as strongly as non affine displacement.

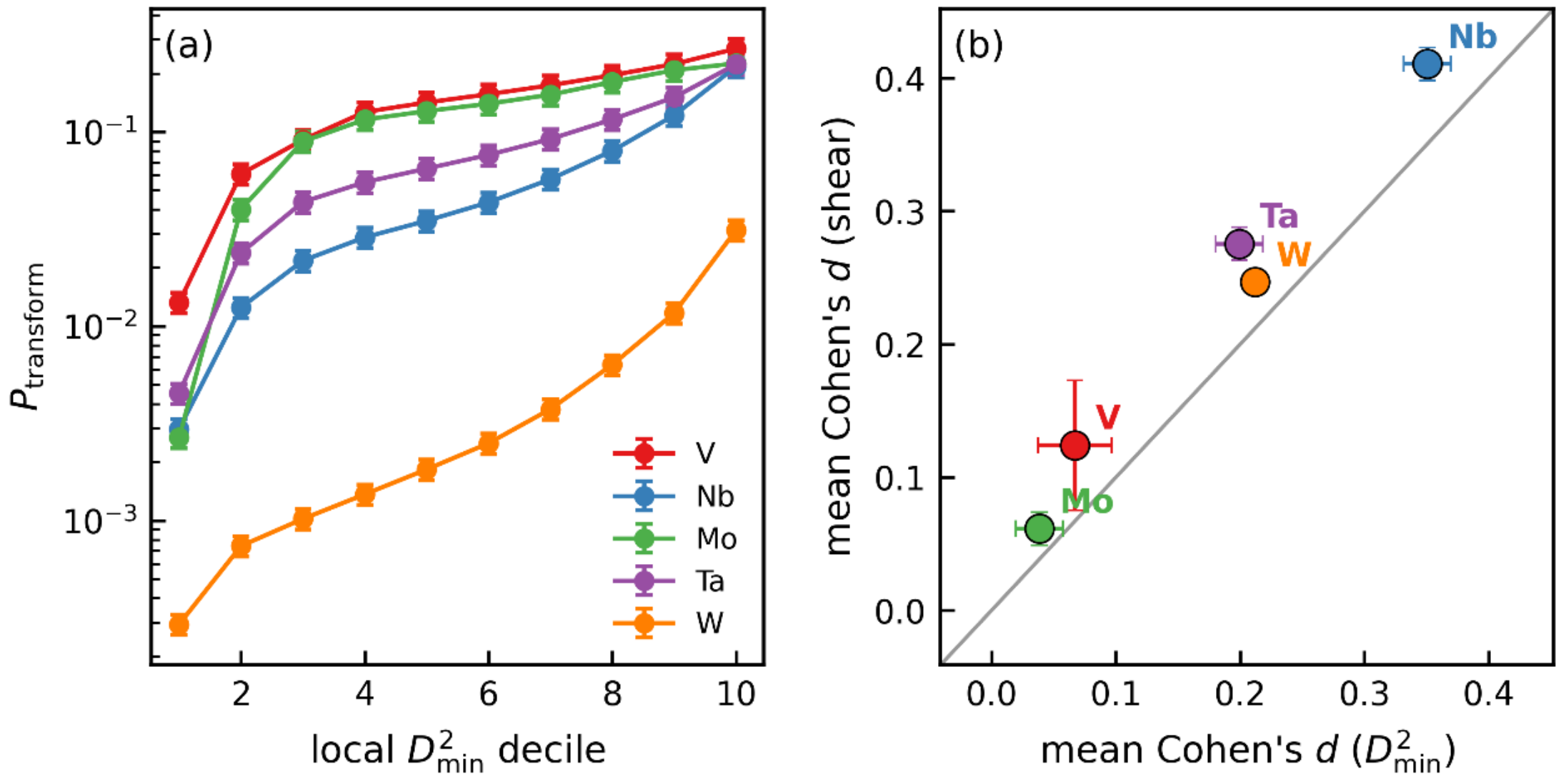

**Figure 8.** Association between local deformation and subsequent persistent transformation. (a) Persistent transformation probability as a function of precursor $D^2_{\text{min}}$ decile. (b) Cohen's $d$ for local shear strain plotted against that for $D^2_{\text{min}}$. Points on or above the diagonal indicate that local shear strain discriminates future transforming atoms at least as strongly as $D^2_{\text{min}}$ in the corresponding univariate comparisons. The depth resolved Cohen's $d$ results for Mo are presented in Fig. S5.

### 3.7. Material descriptors of crystal forming ability

Indentation induced devitrification is treated here as stress assisted bcc nucleation followed by growth and coalescence. For an ideal spherical embryo, the capillarity contribution to the activation barrier can be written as

$$\Delta G^*_{\text{eff}} = \frac{16\pi\gamma_{ac}^3}{3|\Delta g_{\text{eff}}|^2} \tag{5}$$

$$\Delta g_{\text{eff}} = \Delta g_{c-a} - \sigma : \varepsilon^T + \Delta g_{\text{el}} + \Delta g_{\text{def}} \tag{6}$$

Here $\gamma_{ac}$ is the amorphous crystal interfacial free energy, $\Delta g_{c-a}$ is the signed crystal minus amorphous free energy density difference, $\sigma : \varepsilon^T$ is the mechanical work density associated with the transformation strain, and $\Delta g_{\text{el}}$ and $\Delta g_{\text{def}}$ collect elastic and defect energy. All terms in Eq. (6) have units of energy per volume. This expression is a local framework, not a spatially uniform barrier model for the indentation field. For constant indentation velocity v and transformed fraction ξ, the depth domain CFA is related to a time domain rate by

$$\text{CFA} \equiv \max_h \left(\frac{d\xi}{dh}\right) = \frac{1}{v}\max_t \left(\frac{d\xi}{dt}\right) \propto \frac{v_0}{v}\exp\left(-\frac{\Delta G^*_{\text{eff}}}{k_B T}\right) \tag{7}$$

Here $v_0$ is an attempt frequency, $k_B$ the Boltzmann constant and T the absolute temperature. At 300 K and $v = 0.075$ Å $\text{ps}^{-1}$ the five element descriptor analysis is exploratory (Figure 9).

Cohesive energy reproduces the rank order exactly (Spearman $\rho_s = -1.00$) and explains ln(CFA) with $R^2 = 0.81$, consistent with easier rearrangement in the more weakly bound metals.

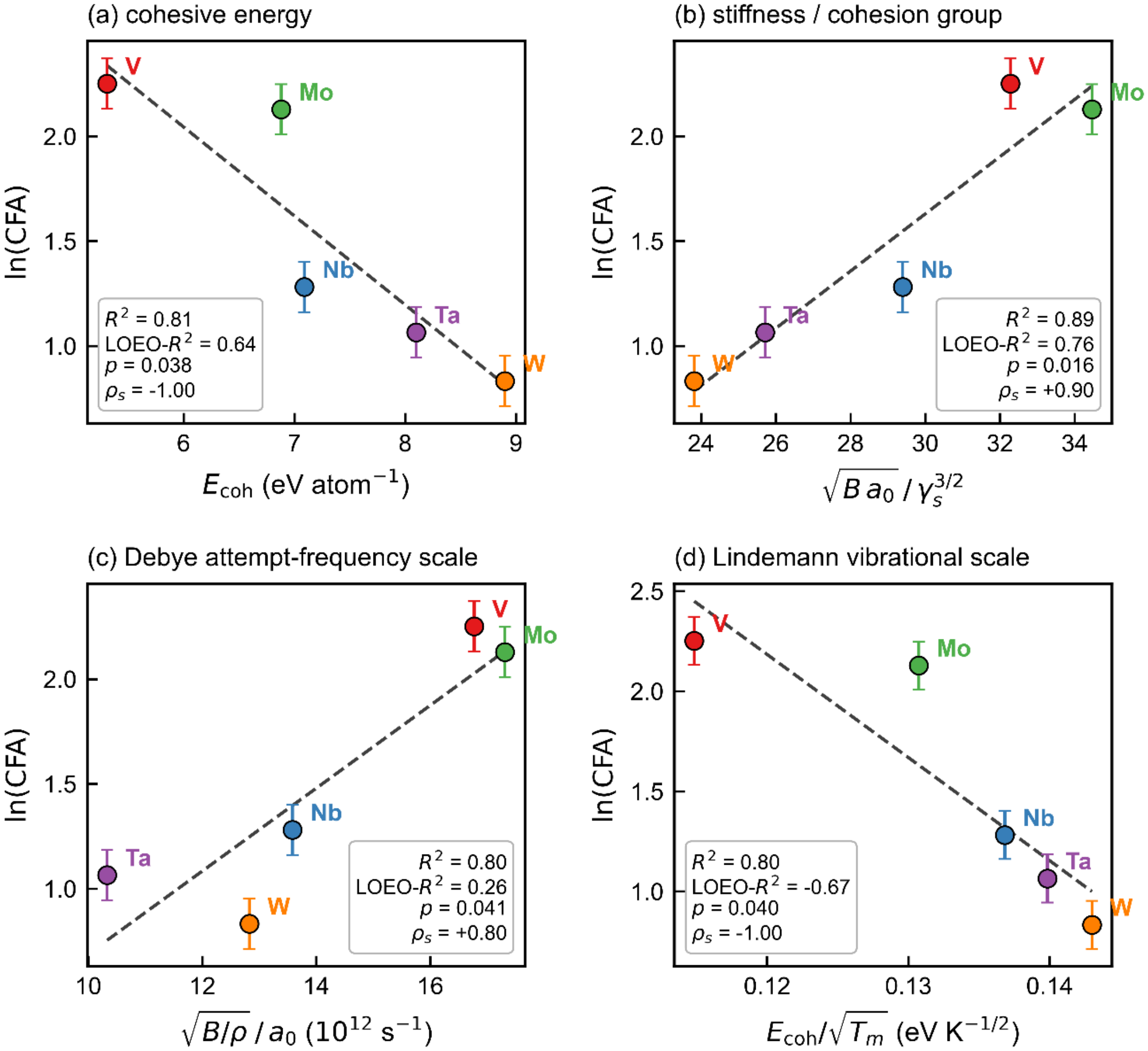

**Figure 9.** Crystal forming ability at the reference condition (300 K, v = 0.075 Å ps$^{-1}$) against four candidate descriptors: (a) cohesive energy $E_{coh}$; (b) the stiffness and cohesion group $\sqrt{(B\,a_0)}/\gamma_s^{3/2}$; (c) the Debye attempt frequency scale $\sqrt{(B/\rho)}/a_0$; (d) the Lindemann vibrational scale $E_{coh}/\sqrt{T_m}$. Each panel reports $R^2$, leave-one-element-out (LOEO) $R^2$, the two sided $p$ value and the Spearman $\rho_s$; dashed lines are ordinary least squares fits to ln(CFA).

The descriptor values, the single-descriptor regressions, their correlations and the velocity models are collected in the SM (Tables S1–S5), where the bulk only nucleation barrier diagnostic and the melting temperature robustness of the Lindemann fit are also shown (Figs S6 and S7).

### 3.8. Early bcc like order correlates with the depth domain kinetics

Across the five element trajectories the first analysed bcc like fraction spans almost two orders of magnitude and correlates with both the effective depth exponent and the half transformation depth (Figure 10). With the early order index defined as $\varphi_{0.1} = f_{\mathrm{bcc}}(0.1\ \mathrm{nm})/(1\%)$, the five-element fits are

$$n_h = \frac{3}{2} + 2.709\varphi_{0.1}^{-0.688}, R^2 = 0.982, Q^2_{\mathrm{LOEO}} = 0.963 \tag{8}$$

$$h_{50} = 2.437\varphi_{0.1}^{-0.656}\mathrm{nm}, R^2 = 0.965, Q^2_{\mathrm{LOEO}} = 0.915 \tag{9}$$

where $Q^2_{\mathrm{LOEO}}$ is the leave one element out predictive coefficient, defined as $Q^2_{\mathrm{LOEO}} = 1 - \Sigma_i(y_i - \hat{y}_{(-i)})^2/\Sigma_i(y_i - \bar{y})^2$, where $y_i$ is the fitted quantity for element i, $\hat{y}_{(-i)}$ its value predicted by the fit to the other four elements and $\bar{y}$ the five-element mean. The fitted crossover to the classical value $n_h = 4$ occurs at $\varphi_{0.1} \approx 1.12$ ($n_h = 4$ is the Avrami exponent for a constant nucleation rate with three-dimensional growth, with $h \propto t$ at constant v). When the baseline $f_0$ is fitted rather than fixed at $x = 0$, the exponents fall to 0.489 and 0.491; the monotonic relation survives, but the numerical exponents are sensitive to the baseline convention (SM, Table S6 and Section S5).

Table 3 collects the early bcc-like fraction $f_{\mathrm{bcc}}(0.1\ \mathrm{nm})$, the fitted depth exponent $n_h$, the half-transformation depth $h_{50}$ and the terminal transformed fraction $f_\infty$ of each element. The terminal fraction ranges from 69.0% in W to 85.4% in Nb and does not follow the CFA order. In Eq. (8) the constant 3/2 is a lower bound of $n_h$ fixed before fitting, not a fitted or measured growth exponent; it was chosen because, for shallow contact of a sphere ($h \ll R$), the contact radius $a \approx (2Rh)^{1/2}$ makes the contact-volume scale $a^3$ grow as $h^{3/2}$ [61]. The prefactor and exponent of Eq. (8) depend on this choice, and Eq. (8) is therefore an empirical description of the five-element series.

**Table 3.** Early structural state and KJMA-like depth descriptors of each element (v = 0.075 Å ps$^{-1}$). For W, $f_\infty$ was fixed to the terminal bcc fraction of an extended trajectory because the transformation is incomplete at h = 10 nm.

| Element | $f_{\mathrm{bcc}}(0.1\ \mathrm{nm})$ (%) | $n_h$ | $h_{50}$ (nm) | $f_\infty$ (%) |
|---|---|---|---|---|
| V | 1.5089 | 4.200 | 1.587 | 78.54 |
| Nb | 0.2102 | 9.717 | 7.874 | 85.37 |
| Ta | 0.2578 | 7.429 | 6.073 | 83.16 |
| Mo | 6.4418 | 2.144 | 0.773 | 77.98 |
| W | 0.1114 | 13.614 | 9.428 | 69.02 |

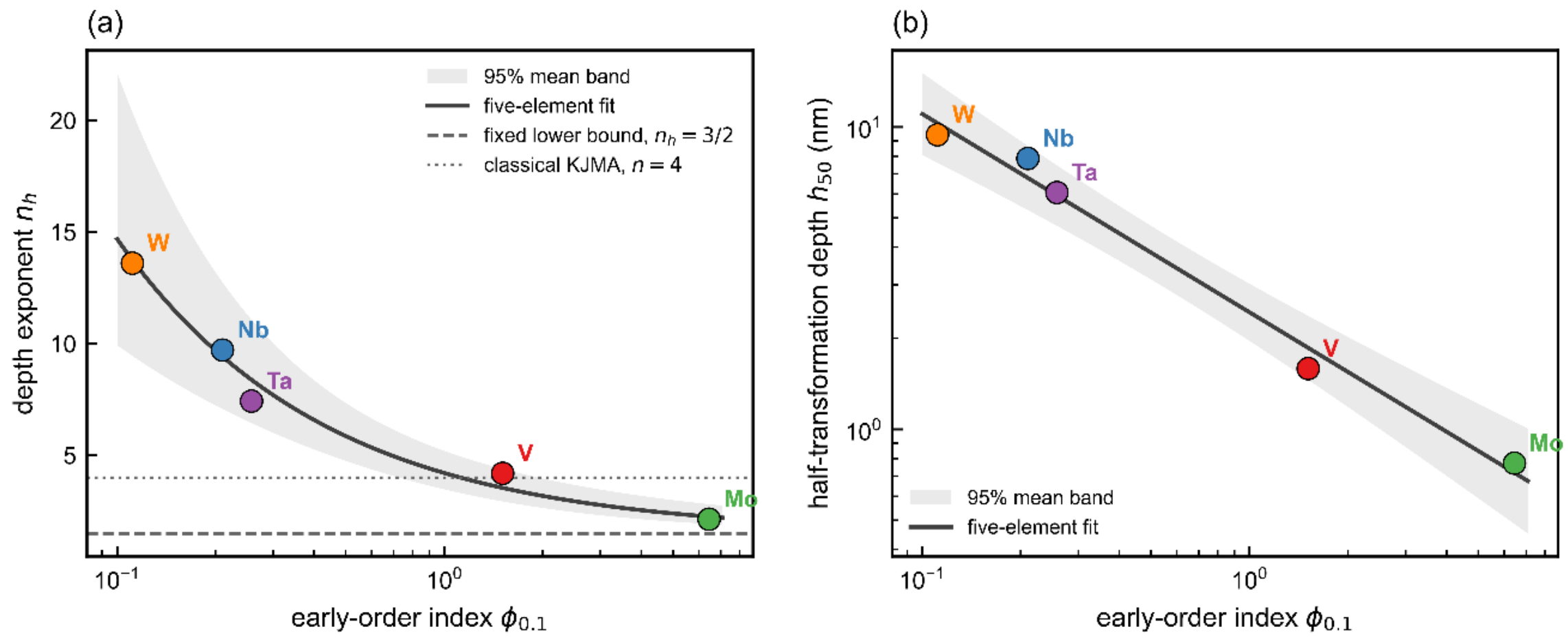

**Figure 10.** Early-order index versus (a) effective depth exponent and (b) half-transformation depth for the five elements. Lines are the fits of Eqs. (8) and (9) and shaded regions are 95% confidence bands of the mean; the dashed line in (a) marks the fixed lower bound $n_h = 3/2$ of Eq. (8).

### 3.9. Work to half transformation and peak cluster density

The half-transformation depth of the KJMA fit is $h_{50} = h_1 + h_c(\ln 2)^{1/n_h}$ (Table 3). The mechanical work done by the indenter up to this depth is $W_{50} = \int_0^{h50} F(h)\, dh$, obtained by trapezoidal integration of the load–depth curve; it ranges from 383 eV (Mo) to $1.8 \times 10^5$ eV (W) and is a macroscopic mechanical work, not an atomistic activation barrier. The peak cluster density is $n_{g,max} = N_{max}/V_A$, where $N_{max}$ is the maximum of the retained-cluster count $N_g(h)$ (SM, Fig. S4) and $V_A$ is the analysed specimen volume (3.34–3.44 × $10^4$ nm$^3$, 95–99% of the slab volume; SM, Table S8). With the potential-specific B and $a_0$ of Table 1, the indenter radius R = 10 nm, the dimensionless work to half transformation and peak cluster density are

$$\Pi_W = \frac{W_{50}}{BR^3}, \qquad \Pi_n = n_{g,\max}\, a_0^3 \qquad (10)$$

The resulting descriptors are listed in Table 4. $\Pi_W$ spans more than two orders of magnitude and orders the series $\mathrm{Mo} < \mathrm{V} < \mathrm{Ta} < \mathrm{Nb} < \mathrm{W}$. The peak cluster density follows

$$\Pi_n = 5.00 \times 10^{-6}\, \Pi_W^{-0.479}, \qquad R^2 = 0.999, \quad Q^2_{LOEO} = 0.994\ (N = 5) \qquad (11)$$

The 95% confidence interval of the exponent $[-0.512, -0.447]$ contains $-1/2$, and the constrained form is $\Pi_n\sqrt{\Pi_W} = 4.52 \times 10^{-6}$. Equation (11) is a compact series specific response law rather than a universal exponent. The dimensionless descriptors are tabulated in the SM (Table S7).

**Table 4.** Dimensionless work to half transformation and peak cluster density. All dimensionless quantities use the potential specific B and $a_0$ of Table 1; $n_{g,max} = N_{max}/V_A$.

| Element | $\Pi_W$ | $n_{g,\max}$ (nm$^{-3}$) | $\Pi_n$ | $N_{\max}$ |
|---|---|---|---|---|
| V | $8.061 \times 10^{-4}$ | $5.874 \times 10^{-3}$ | $1.634 \times 10^{-4}$ | 196 |
| Mo | $2.023 \times 10^{-4}$ | $8.912 \times 10^{-3}$ | $2.783 \times 10^{-4}$ | 299 |
| Ta | $3.184 \times 10^{-2}$ | $7.295 \times 10^{-4}$ | $2.631 \times 10^{-5}$ | 25 |
| Nb | $5.044 \times 10^{-2}$ | $5.818 \times 10^{-4}$ | $2.106 \times 10^{-5}$ | 20 |
| W | $9.037 \times 10^{-2}$ | $4.967 \times 10^{-4}$ | $1.538 \times 10^{-5}$ | 17 |

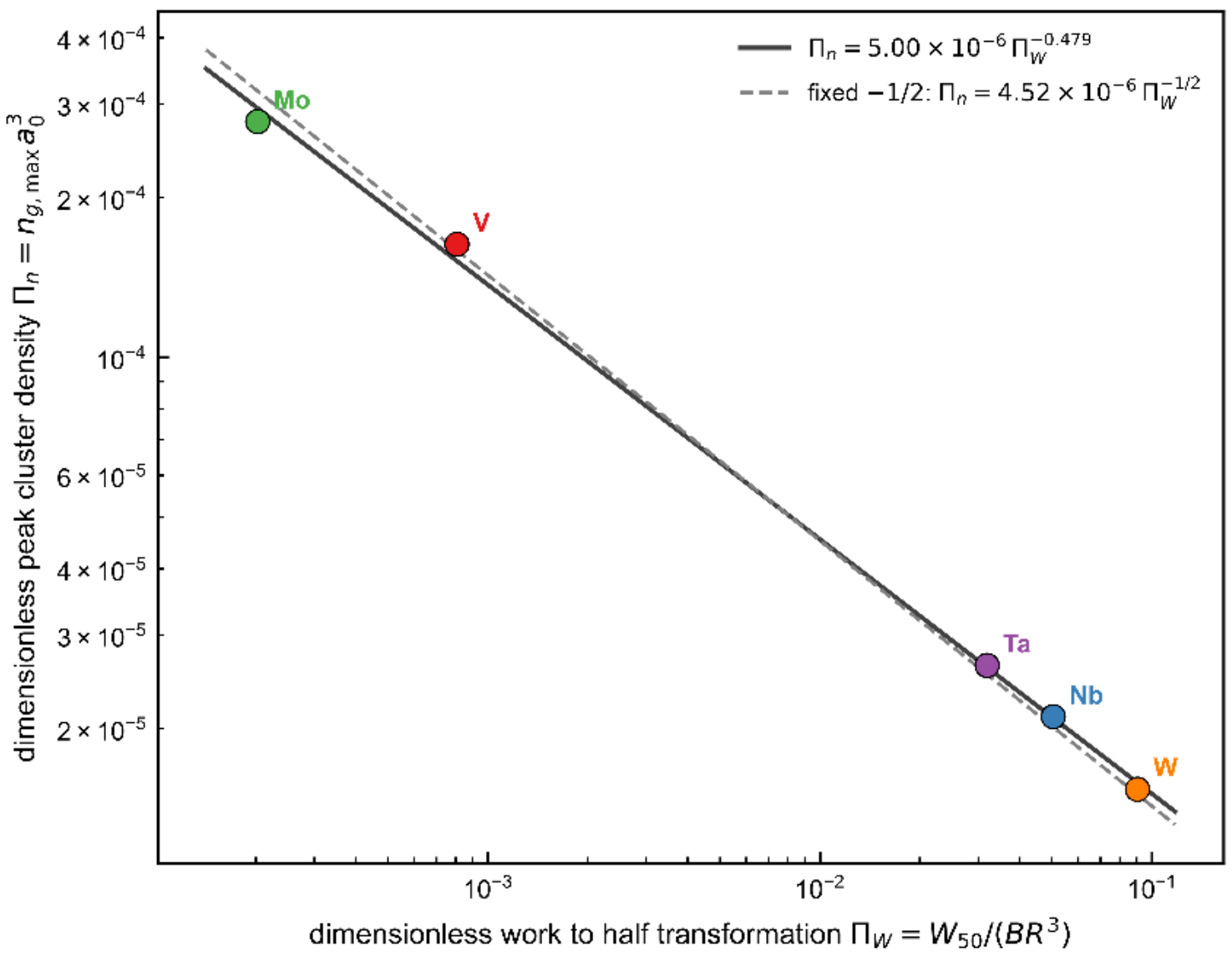

**Figure 11.** Dimensionless peak cluster density $\Pi_n$ versus dimensionless work to half transformation $\Pi_W$ for the five elements; the solid line is Eq. (11) and the dashed line the constrained form with exponent −1/2.

### 3.10. Nucleation, growth and coalescence population balance

The crystalline volume carried by persistent bcc clusters is

$$V_{\mathrm{BCC}} = N_g \bar{V}_g \approx f_{\mathrm{bcc}} V_A \tag{12}$$

where $N_g$ is the number of retained bcc clusters, $\bar{V}_g$ their mean volume and $V_A$ the analysis volume. The exact midpoint identity

$$\Delta(NV) = V_{\mathrm{mid}} \Delta N + N_{\mathrm{mid}} \Delta V \tag{13}$$

separates the volume increment associated with a change in cluster number from that associated with growth of the existing population. In Eq. (13), $\Delta N$ and $\Delta V$ are the increments of $N_g$ and $V_g$ between successive saved configurations, and $N_{mid}$ and $V_{mid}$ their averages over the same interval. Growth supplies 92.8% of the pre peak crystalline volume increase in V, 79.7% in Mo, 63.9% in Ta, 68.6% in Nb and 82.0% in the W analysis (Table 5; SM, Fig. S8). In the final analysed configuration (subscript f) the equivalent spherical diameter of the mean cluster is $d_{eq,f} = (6\bar{V}_{g,f}/\pi)^{1/3}$, where $\bar{V}_{g,f}$ is the total volume of the retained clusters divided by their number $N_f$; it is a volume-based size and in general exceeds the arithmetic mean of the individual cluster diameters. Combined with Eq. (12), this definition gives

$$\Pi_d^3\,(1-C)\,\Pi_n = \frac{6\,f_{\text{terminal}}}{\pi}, \qquad \Pi_d = \frac{d_{\text{eq},f}}{a_0}, \qquad C = 1 - \frac{N_f}{N_{\max}} \tag{14}$$

where $C$ is the net fraction of the peak cluster count $N_{\max}$ lost by impingement and coalescence, and $f_{terminal} = N_f\bar{V}_{g,f}/V_A$ is the fraction of the analysis volume occupied by the retained clusters in the final configuration, close to $f_\infty$ of Table 3. Eq. (14) is an exact volume balance rather than an independent prediction; the tabulated values satisfy it within 0.04%, which confirms the internal consistency of the cluster statistics, and its use is to decompose the terminal cluster size into cluster density, coalescence and transformed fraction. Mo and V combine dense persistent populations with strong coalescence, whereas Ta and Nb form fewer clusters that grow to larger terminal sizes of 15.0 and 17.2 nm (Table 5; SM, Fig. S9).

**Table 5.** Pre peak population balance partition and terminal coalescence descriptors. $C = 1 - N_f/N_{max}$; $N_f$ is the number of retained clusters and $d_{eq,f}$ the equivalent spherical diameter of the mean cluster in the final analysed configuration. For W, whose transformation is incomplete at h = 10 nm, $N_f$ is the completion estimate $f_{terminal}V_A/\bar{V}_{g,f}$ based on the extended trajectory.

| Element | Number term (%) | Growth term (%) | $C$ | $N_f$ | $d_{eq,f}$ (nm) |
|---|---|---|---|---|---|
| V | 7.2 | 92.8 | 0.663 | 66 | 9.16 |
| Mo | 20.3 | 79.7 | 0.756 | 73 | 8.86 |
| Ta | 36.1 | 63.9 | 0.360 | 16 | 15.04 |
| Nb | 31.4 | 68.6 | 0.450 | 11 | 17.24 |
| W | 18.0 | 82.0 | 0.216 | 13.33 | 15.01 |

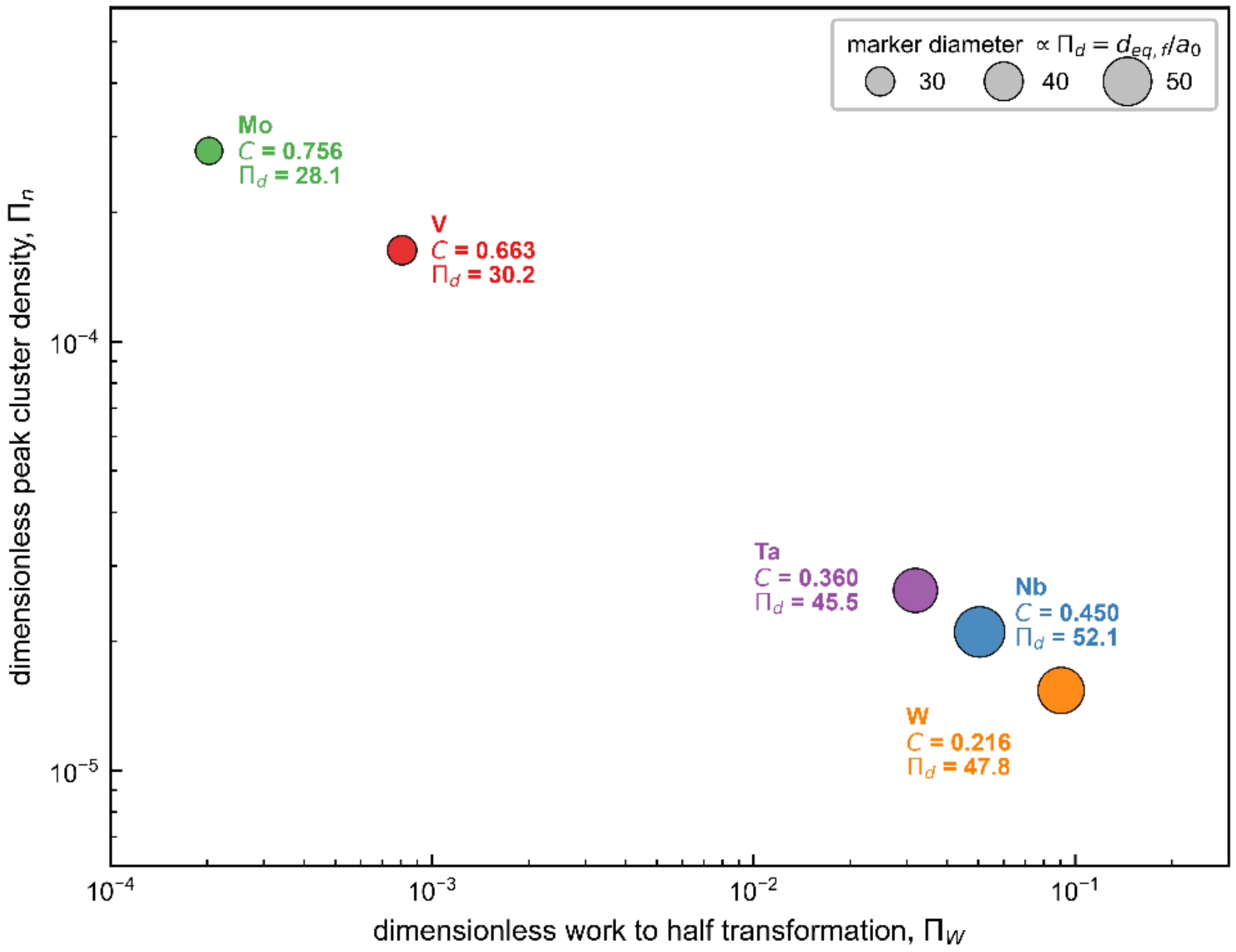

**Figure 12.** Mechanochemical pathway map: dimensionless peak cluster density $\Pi_n$ versus dimensionless work to half transformation $\Pi_W$. Marker diameter is proportional to $\Pi_d = d_{eq,f}/a_0$ (size key in the legend); labels give the coalescence fraction C and $\Pi_d$.

Figure 12 assembles these descriptors into a mechanochemical pathway map that separates the five elements by their nucleation, growth and coalescence behaviour, with the marker diameter proportional to $\Pi_d$ and the labels reporting the coalescence fraction C.

## 4. Discussion

At the pair correlation level, the selected interatomic potentials reproduce the broad liquid structure features of the ab initio references, including the first-neighbour maximum, the first minimum and the second-neighbour maximum, and the classical melts are diffusive liquids with self-diffusion coefficients of the same order as the AIMD values. This agreement supports their use for a comparative study of melt quenching and indentation induced ordering, but it does not establish quantitative fidelity of the amorphous free energies or of the transformation kinetics. The element-specific reference densities, together with first-peak height deviations of up to 14% (Nb, Mo) and 20% (W), remain relevant limitations, and the comparison should be read as a benchmark of nearest neighbour length scale rather than of thermodynamic accuracy.

Under the common simulation protocol, all five amorphous metals transform locally towards bcc order during indentation. The operational CFA ranks the response as V > Mo > Nb > Ta > W at the reference velocity. Because CFA is the maximum change in bcc classified fraction per unit indenter advance, it combines the intrinsic structural response with the imposed loading rate; it is not an intrinsic nucleation rate, a hardness, or the inverse of glass forming ability. Across the three tested velocities the element specific exponents are approximately inverse, with a mean of 1.08 and a between element standard deviation of 0.08. This dispersion is descriptive of the five element set and should not be read as uncertainty on a universal exponent.

Because the exponents are close to unity, the product CFA·v is nearly constant, so the transformation rate expressed per unit time is almost independent of indenter velocity over the range examined. The depth domain ranking therefore reflects the kinematic conversion between depth and elapsed time as much as any intrinsic rate difference between elements, and the departures from unity, 0.97 for Mo up to 1.18 for Nb, are the part of the response not accounted for by that conversion. We regard this near invariance of CFA·v as the most transferable single observation of the study, because it isolates the kinematic content of the depth domain metric from the material content.

A time lagged, matched case control comparison links the transformation to the local deformation state. Atoms that subsequently satisfy the persistent bcc criterion carry larger precursor non affine displacement and local shear strain than matched atoms that remain amorphous, whereas hydrostatic pressure shows only a small negative contrast (partly by design, since the controls are matched in pressure) and atomic volume is consistently smaller among future transforming atoms. These univariate associations are consistent with a mechanically assisted, shear biased rearrangement. They do not, however, identify an independent causal variable or a unique critical shear stress: only the diagonal stress was stored, so the pressure channel is hydrostatic and the comparison speaks to shear strain rather than to a resolved deviatoric stress. The claim is therefore that transformation is associated with excess local shear strain and non affine displacement, not that a specific stress invariant has been shown to control it.

Two candidate controls on the material ranking can be separated. Cohesive energy reproduces the CFA rank order exactly ($\rho_s = -1.00$) and explains ln(CFA) with $R^2 = 0.81$, whereas a bulk only classical nucleation diagnostic constructed from $\gamma_s$ and $\Delta g_bulk$ has essentially no explanatory power ($R^2 \approx 0.00$), and the estimated bulk driving force is in fact largest for W, the most resistant element in the series. Resistance to indentation induced devitrification therefore tracks cohesive bond strength rather than the thermodynamic driving force, consistent with easier local rearrangement in the more weakly bound metals. This does not render the bulk free energy difference irrelevant: the diagnostic omits the local stress work, the transformation strain, the amorphous crystal interfacial energy and the atomic mobility, all of which enter the true barrier. The candidate descriptors are moreover strongly collinear across the five elements, so these single descriptor regressions identify correlations but cannot isolate a unique controlling property, and the rank correlation should be read as a property of this potential set rather than as an established material law.

The early bcc like order and the mechanical work descriptor summarise different aspects of the response. Across the five elements the early order index $\varphi_{0.1}$ is strongly associated with both the fitted depth response sharpness and the half transformation depth, and the positive leave one element out $Q^2$ values indicate internal consistency within this five point series rather than external predictive validation. The fitted early order exponents also depend on the baseline convention: fixing the baseline at the first analysed depth and fitting it freely give different numerical exponents, although the monotonic relation survives in both cases. The normalised work analysis covers all five elements and orders the dimensionless work to half transformation as Mo < V < Ta < Nb < W. The log–log fit gives $\Pi n = 5.00 \times 10^{-6}\ \Pi W^{(-0.479)}$ with $R^2 = 0.999$ and $Q^2_{LOEO} = 0.994$; the fitted exponent is statistically consistent with −½, and imposing that value gives the compact form $\Pi n\ \Pi W^{(1/2)} = 4.52 \times 10^{-6}$. Because all five elements define the relation and the descriptors are collinear, the near square root dependence is reported as an empirical regularity of this series rather than as a nucleation law derived from first principles.

After persistent clusters form, the population balance shows that grain growth contributes between 63.9% (Ta) and 92.8% (V) of the pre peak transformed volume increase, and the exact volume balance of Eq. (14), satisfied within 0.04% for every element, decomposes the terminal cluster size into peak cluster density, coalescence and transformed fraction. The balance provides an internally consistent accounting of changes in cluster number, cluster growth and coalescence, and shows that the terminal microstructure is set jointly by how many nuclei persist and how strongly they coalesce, rather than by the CFA ranking alone: Mo and V combine dense persistent populations with strong coalescence, whereas Ta and Nb form fewer clusters that grow to larger terminal sizes.

Several limitations bound this interpretation. The glasses are monatomic, potential dependent configurations produced by ultrafast quenching, and every indentation velocity examined lies far above experimental rates; the lowest velocity used here is itself set by the accessible simulation time, so this rate gap is an intrinsic constraint of the method rather than a parameter that could be reduced within the present framework. The five element scaling relations are evaluated on ensemble mean descriptors, each averaged over three independent glass replicas and ten seeded indentation trajectories per element, so they are not sensitive to the selection of any individual run; they nonetheless rest on five element level points and remain sensitive to the choice of interatomic potential, to the polyhedral template matching and cluster persistence definitions, and to the fixed boundary, cell geometry and thermostatting. The specimen is a finite slab the maximum analysed depth (≈10 nm) is ≈0.46 of the slab thickness H ≈ 21 nm, far beyond the ≈10% of film thickness below which substrate effects are usually neglected in film indentation [62], and at the largest depths the transformed region reaches the lateral periodic boundaries and approaches the fixed layer (Fig. 7). The terminal fraction, the cluster counts $N_{max}$ and $N_f$, the coalescence fraction C and the late stage of $f_{bcc}(h)$ are therefore properties of the common 41 × 41 × 21 $nm^3$ cell, and because the inflection depths of Nb, Ta and W (≈6-9 nm) fall where the stress field interacts with the fixed layer, a cell-size dependence of their CFA cannot be excluded. The comparison between elements remains meaningful because all five were simulated in the same geometry. The stored stress data are hydrostatic and do not support a fully resolved deviatoric analysis. The central contribution is therefore a set of testable, series specific links among early bcc like order, mechanically associated transformation and grain population evolution, rather than universally calibrated laws. The robustness of these links to interatomic potential, quench rate, tip radius and cell size is being examined systematically in separate, ongoing work and lies beyond the scope of the present study.

## 5. Conclusions

Large-scale molecular dynamics shows a localised, persistent amorphous to bcc transformation in melt quenched V, Nb, Mo, Ta and W under spherical nanoindentation at 300 K, proceeding by bulk nucleation, growth and coalescence; a time matched calculation without an indenter shows no such transformation, indicating that the ordering is associated with contact loading rather than with thermal evolution over the same interval.

The operational crystal forming ability decreases as V > Mo > Nb > Ta > W and shows an approximately inverse dependence on indenter velocity over 0.075 to 0.30 Å $ps^{-1}$, so that CFA·v is nearly constant and the time domain transformation rate is almost velocity independent. Atoms that subsequently transform carry larger preceding non affine displacement and local shear strain than matched amorphous controls, consistent with a shear biased local rearrangement rather than proof of a unique causal stress variable. Resistance to transformation tracks cohesive bond strength rather than the thermodynamic driving force, which is in fact largest for the most resistant element, W; a bulk only classical nucleation diagnostic does not reproduce the ranking.

Across the five elements, the ensemble mean early bcc like fraction correlates with both the depth response sharpness and the half transformation depth, and the normalised peak cluster density decreases with the dimensionless work to half transformation with a fitted exponent of −0.479, statistically consistent with −½, at an internal leave one element out $Q^2$ of 0.994. Because all five elements define this relation, it is an internally assessed series correlation rather than an external validation or a universal scaling law. Grain growth accounts for 64 to 93% of the pre peak transformed volume increase, while the algebraic population balance relates peak cluster density and coalescence to the terminal equivalent cluster size.

Together these results provide a common, normalised basis for comparing stress driven devitrification across the bcc refractory metals, and connect early structure, mechanical loading and microstructure evolution during mechanically induced devitrification within this five element, potential specific series.

**Acknowledgements**

This work was supported by the Ministry of Education, Youth and Sports of the Czech Republic through e-INFRA CZ (ID:90254). The computing part was partially co-funded by the European Union under the project Robotics and advanced industrial production (reg. no. CZ.02.01.01/00/22_008/0004590).